\documentclass{article}

\usepackage{microtype}
\usepackage{graphicx}
\usepackage{subfigure}
\usepackage{booktabs} 
\usepackage[T1]{fontenc}
\usepackage{hyperref}

\usepackage[accepted]{icml2025}

\usepackage{fp} 
\usepackage{hyperref}
\usepackage{url}
\usepackage{graphicx}
\usepackage{colortbl}
\usepackage{color}
\usepackage{siunitx} 
\definecolor{firstblue}{RGB}{66,146,198}
\definecolor{secondblue}{RGB}{158,202,225}
\definecolor{thirdblue}{RGB}{222,235,247}
\definecolor{go in new uniref}{RGB}{218,168,124}
\definecolor{go supprot by family}{RGB}{124,152,149}
\definecolor{transmembrane}{RGB}{0,142,231}
\definecolor{none-transmembrane}{RGB}{1,238,2}
\definecolor{decay}{RGB}{169,169,169}
\definecolor{overall}{RGB}{247,104,161}
\usepackage{amsmath}
\usepackage{amssymb}
\usepackage{mathtools}
\usepackage{amsthm}
\usepackage{amsfonts}       
\usepackage{nicefrac}       
\usepackage{appendix}
\usepackage{makecell}
\usepackage{newfloat}
\usepackage{multirow}
\usepackage{booktabs}       
\usepackage{listings}
\usepackage{authblk}
\usepackage{wrapfig}
\usepackage{pifont}
\usepackage{algorithm}
\usepackage{algorithmic}
\usepackage{bbm}
\usepackage{tabularx}
\usepackage[capitalize,noabbrev]{cleveref}

\theoremstyle{plain}

\theoremstyle{definition}

\theoremstyle{remark}

\usepackage[textsize=tiny]{todonotes}

\icmltitlerunning{Submission and Formatting Instructions for ICML 2025}

\begin{document}

\twocolumn[
\icmltitle{AnnoDPO: Protein Functional Annotation Learning\\ with Direct Preference Optimization}




\begin{icmlauthorlist}
\icmlauthor{Zixuan Jiang}{yyy}
\icmlauthor{Renjing Xu}{yyy}
\end{icmlauthorlist}

\icmlaffiliation{yyy}{The Hong Kong University of Science and Technology (Guangzhou)}

\icmlcorrespondingauthor{Renjing Xu}{renjingxu@hkust-gz.edu.cn}

\icmlkeywords{Machine Learning, ICML}

\vskip 0.3in
]



\printAffiliationsAndNotice{}  

\begin{abstract}
Deciphering protein function remains a fundamental challenge in protein representation learning. The task presents significant difficulties for protein language models (PLMs) due to the sheer volume of functional annotation categories and the highly imbalanced distribution of annotated instances across biological ontologies. Inspired by the remarkable success of reinforcement learning from human feedback (RLHF) in large language model (LLM) alignment, we propose AnnoDPO, a novel multi-modal framework for protein function prediction that leverages Direct Preference Optimization (DPO) to enhance annotation learning. Our methodology addresses the dual challenges of annotation scarcity and category imbalance through preference-aligned training objectives, establishing a new paradigm for biological knowledge integration in protein representation learning. We provide the code for AnnoDPO at \url{https://github.com/AzusaXuan/AnnoDPO}.
\end{abstract}

\begin{figure*}[htb]
    \centering
    \includegraphics[width=1\linewidth]{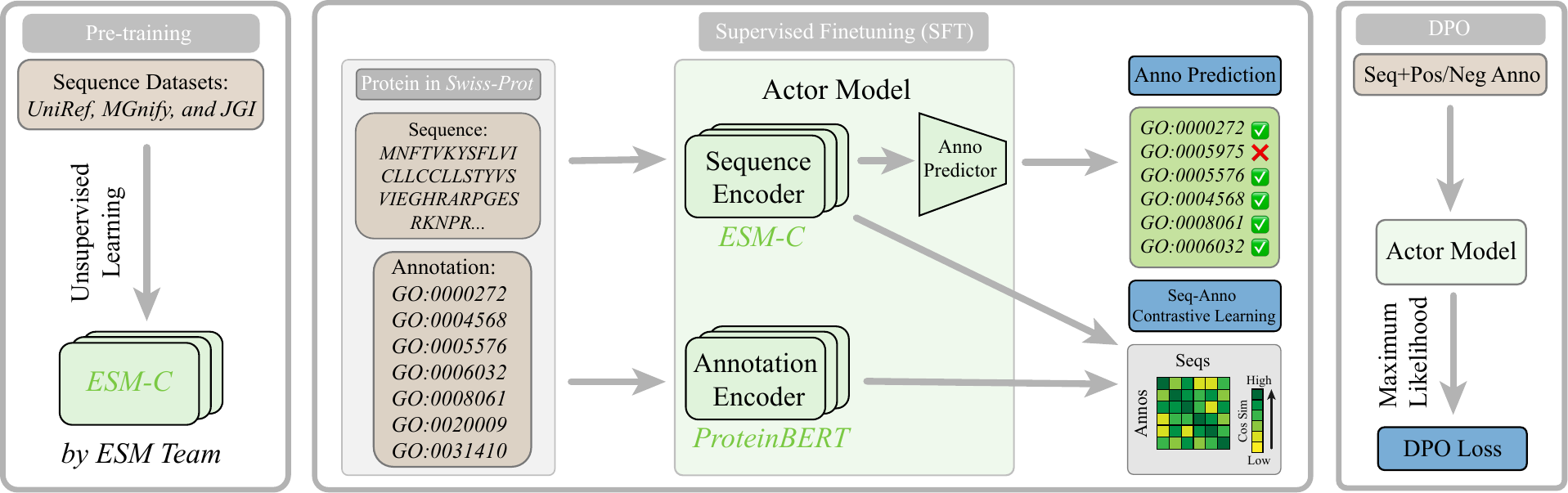}
    
    \caption{Model architecture and training objectives of AnnoDPO. The training framework is divided into three stages: \textbf{Pre-training}: Self-supervised learning of ESM-C on protein sequences from UniRef, MGnify, and JGI \cite{esmteam2024esmc}; \textbf{SFT}: Dual-objective finetuning with annotation prediction and sequence-annotation contrastive alignment; \textbf{DPO}: Preference optimization through positive annotations against negative ones.}
    
    \label{fig:Method}
\end{figure*}

\section{Introduction}
Proteins serve as the central machinery of life, executing crucial biological activities. While high-throughput sequencing technologies \cite{reuter2015high} have driven exponential growth in sequenced genomes over two decades \cite{uniprot2019uniprot,suzek2015uniref}, functionally characterized proteins \cite{boeckmann2003swiss,gasteiger2001swiss} lag significantly due to structural complexity and challenges in capturing interaction dynamics. This disparity underscores the persistent challenge of accurate, large-scale automated protein function prediction \cite{radivojac2013large,friedberg2006automated}.

Traditional approaches for functional annotation—including statistical methods and rule-based systems like UniRule—remain widely adopted in protein databases \cite{uniprot2019uniprot,dougan2016uniprot,sledz2016structural}. However, their reliance on simplified sequence-function mappings often leads to inaccuracies. Deep learning methods \cite{kulmanov2018deepgo,you2021deepgraphgo,kulmanov2020deepgoplus,kulmanov2024protein,yu2023enzyme,jang2024accurate} have recently emerged as superior alternatives, with PLMs \cite{elnaggar2021prottrans,brandes2022proteinbert,rives2021biological,meier2021language} revolutionizing prediction capabilities. However, PLMs face two fundamental challenges: discerning subtle sequence variations that induce dramatic functional divergence and overcoming extreme annotation sparsity where fewer than 5\% of Swiss-Prot entries contain more than 10 Gene Ontology annotations. These combined limitations maintain a persistent accuracy gap between computational predictions and expert annotations, underscoring the need to integrate domain knowledge into PLM-guided functional inference.

A crucial breakthrough has emerged in LLM alignment through RLHF \cite{christiano2017deep, ziegler2019fine, ouyang2022training, bai2022training, glaese2022improving}, which enables AI systems to better align with human preferences. Building on these successes in natural language processing, researchers have begun exploring RLHF's potential for protein-related AI applications. Recent demonstrations span controllable protein generation \cite{liu2025controllable, stocco2024guiding, widatalla2024aligning} and protein knowledge assistants \cite{zhou2025decoding}, establishing RLHF as a viable paradigm for biological sequence modeling. Notably, prior work has not yet explored DPO \cite{rafailov2023direct}, a prominent RLHF variant that eliminates reward modeling through direct policy optimization, for protein function annotation prediction.

This study establishes three key contributions: (1) We develop an end-to-end multimodal framework integrating protein sequences with functional annotations, enhanced by contrastive learning during supervised fine-tuning (SFT) to optimize cross-modal feature alignment. (2) We pioneer the adaptation of DPO to protein language models, creating the first DPO-powered architecture for enhancing functional annotation accuracy. (3) We systematically characterize how DPO reshapes model attention patterns to better capture hierarchical relationships in Gene Ontology annotations.

\section{Background}
\paragraph{Protein Functional Annotation Prediction}
Gene Ontology (GO) \cite{ashburner2000gene} provides standardized functional descriptors across three biological domains. Predicting GO terms remains essential for characterizing unannotated proteins. The Enzyme Commission (EC) system \cite{tipton2000history} classifies enzymes via four-digit catalytic activity codes, while UniProtKB keywords (KW) \cite{magrane2011uniprot} systematically categorize functional attributes in Swiss-Prot entries. Together, these annotation systems enable comprehensive protein function analysis.

\paragraph{Protein Multi-modal Learning in Annotation Prediction}
The integration of PLMs with multi-source data has established multimodal learning as the standard for functional annotation. Key advances include: CLEAN \cite{yu2023enzyme} aligning enzymes with EC numbers via contrastive learning; ProteinBERT \cite{brandes2022proteinbert} jointly modeling sequences and GO terms; OntoProtein \cite{zhang2022ontoprotein} encoding knowledge graphs with textual descriptors. Generation paradigms like ProGen \cite{madani2020progen} utilize function labels for controllable synthesis, while ProtST \cite{xu2023protst} bridges sequences with biomedical texts. Most notably, SaProt \cite{su2023saprot} achieves SOTA performance through structure-aware tokenization integrating sequence-structure relationships.

\paragraph{Reinforcement Learning from Human Feedback}
RLHF methodologies bifurcate into reward-modeling and direct preference optimization paradigms. Reward-based approaches \cite{stiennon2020learning, ouyang2022training, christiano2017deep, havrilla2024glore, setlur2024rewarding} employ two-stage training: first learning reward functions from preference data, then optimizing policies via online RL algorithms like PPO \cite{schulman2017proximal}. Conversely, reward-free methods \cite{yuan2023rrhf, song2024preference, dong2023raft} bypass explicit reward modeling by directly optimizing language models on preference rankings. Notably, Direct Preference Optimization (DPO) \cite{rafailov2023direct} has emerged as a predominant reward-free approach due to its stable single-stage training and competitive performance. The field continues to debate fundamental trade-offs: reward-based methods' alignment precision versus reward-free approaches' computational efficiency \cite{li2023policy, xu2024dpo}.

\section{Method}
\begin{figure*}[htb]
    \centering
    \includegraphics[width=1\linewidth]{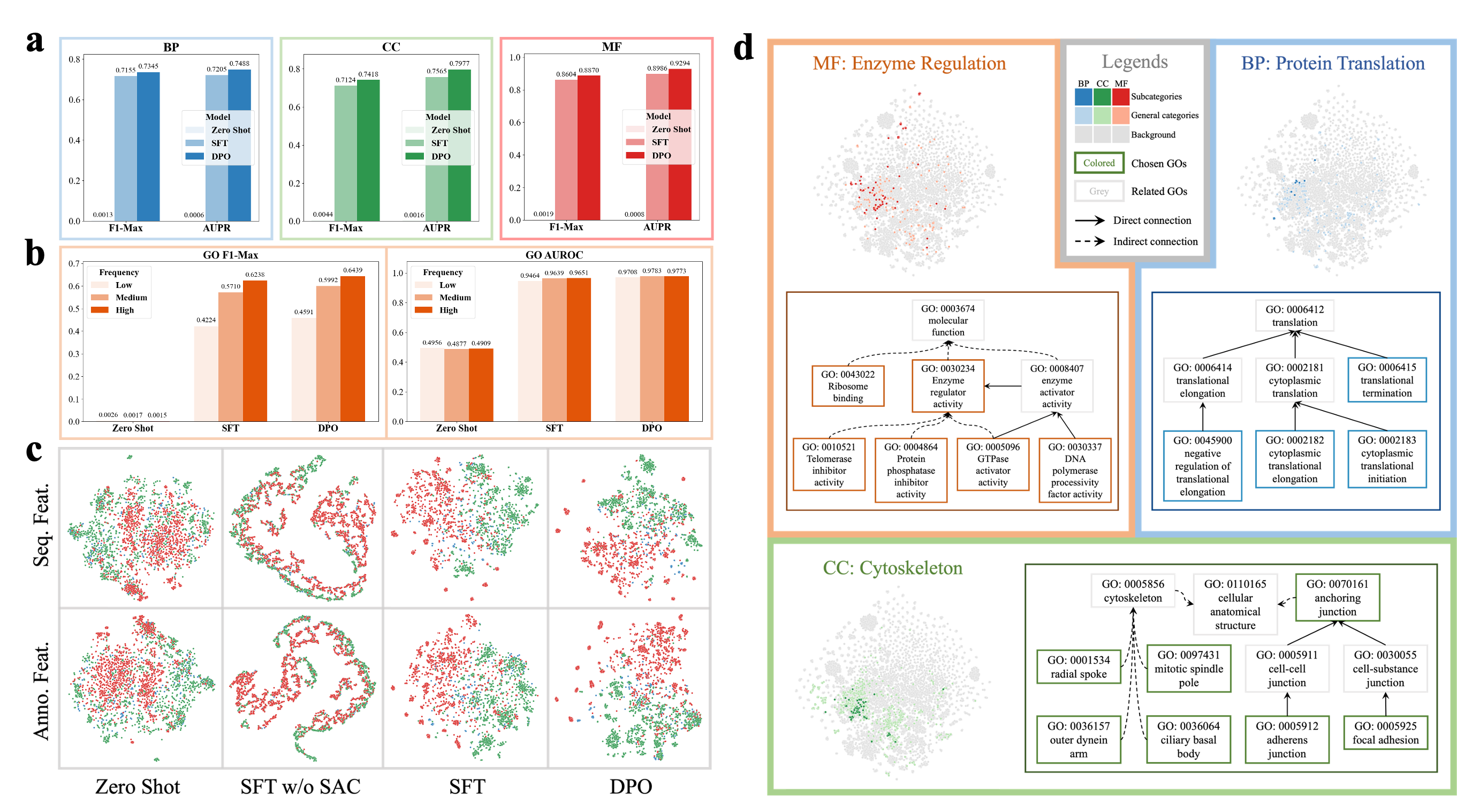}
    
    \caption{Comprehensive Evaluation of Protein Function Annotation Performance.
    \textbf{(a)} Cross-category performance comparison (numerical results in Tab. \ref{tab:cross-category-metrics}).
    \textbf{(b)} Robustness analysis across label frequency regimes (numerical results in Tab. \ref{tab:frequency-robustness}).
    \textbf{(c)} t-SNE visualization of GO category discriminability in latent space.
    \textbf{(d)} Hierarchical relationship preservation in tightly-related GO term families (additional examples in Appendix \ref{add_exp_results}).}
    \label{fig:go_exampe_main}
\end{figure*}
Our three-stage training framework (Fig. \ref{fig:Method}) comprises pre-training, supervised finetuning (SFT) with combined annotation prediction and sequence-annotation contrastive objectives, and Direct Preference Optimization (DPO). The pre-training stage builds upon ESM Cambrian (ESM-C) \cite{esmteam2024esmc}, where we employ the 300M parameter variant as our foundational sequence encoder. We elaborate the details of SFT and DPO in the subsequent sections and hyperparameter details in Appendix \ref{model_train_details}.

\paragraph{Dataset Curation and data input}
We use Swiss-Prot \cite{boeckmann2003swiss} as the training set as it is one of the most widely used dataset for protein function. To ensure enough sequences for test, we choose the dataset version updated in Jan. 2010 totaling $\sim$510,000 sequences and spilit it at the ratio 9:1 for training and testing. Then we select all the sequences updated after that to construct the Swiss-Prot-New dataset totaling $\sim$60,000 sequences. We demonstrate dataset details in Appendix \ref{dataset_details}.

\paragraph{Supervised Finetuning (SFT)}
The SFT stage integrates three core components: (1) a pretrained ESM-C sequence encoder \cite{esmteam2024esmc} that converts protein sequences into embeddings, (2) an MLP-based annotation predictor generating GO term probabilities from sequence embeddings, and (3) a de novo trained ProteinBERT annotation encoder \cite{brandes2022proteinbert} that encodes functional annotations. We establish cross-modal alignment through contrastive learning between sequence embeddings and annotation features via the sequence-annotation contrastive loss, while simultaneously optimizing annotation prediction accuracy through standard classification objectives. The mathematical formulations of these dual losses are defined as follows:

\paragraph{Annotation Prediction (AP) Loss}
This loss is a sum of the categorical cross-entropy over the protein sequences and the binary cross-entropy over the annotations, namely
\begin{equation}
    \mathcal{L}_{\text{AP}}=-\sum_{j\in N}\left(y^A_j\operatorname{log}(p^A_j)+(1-y^A_j)\operatorname{log}(1-p^A_j)\right),
    \label{ap}
\end{equation}
where $N=7533$ denotes the size of our curated Gene Ontology vocabulary, $y^A_{j} \in \{0, 1\}$ indicates the presence of the  $j$-th GO term in the ground-truth annotations, and $p^{A}_{j} \in [0, 1]$ represents the predicted probability for that term. The GO vocabulary was constructed by retaining terms with over 100 times occurrences in Swiss-Prot, ensuring sufficient statistical support for reliable learning.

\paragraph{Sequence-Annotation Contrastive (SAC) Loss}
$\mathcal{L}_{\mathrm{SAC}}$  implements bidirectional alignment between sequence features $\mathrm{h}^{S}$ and annotation features $\mathrm{h}^{A}$ through normalized feature matching. Given a positive pair $(\mathrm{h}^{S},\mathrm{h}^{A})$ where 
$i$ indexes protein sequences and $j$ indexes functional annotations, the loss computes symmetrized similarity distributions over negative samples:
\begin{equation}
\resizebox{1.0\linewidth}{!}{$\displaystyle
\mathcal{L}_{SAC} = -\frac{1}{2}\sum_{(i,j)}
\left( 
\log\frac{\exp\left(\frac{\mathrm{h}_i^{S} \cdot \mathrm{h}_j^{A}}{\tau}\right)}{\sum_k \exp\left(\frac{\mathrm{h}_i^{S} \cdot \mathrm{h}_k^{A}}{\tau}\right)}
+ \log\frac{\exp\left(\frac{\mathrm{h}_i^{S} \cdot \mathrm{h}_j^{A}}{\tau}\right)}{\sum_k \exp\left(\frac{\mathrm{h}_k^{S} \cdot \mathrm{h}_j^{A}}{\tau}\right)}
\right).
$}
\end{equation}
Here $\tau$ is the temperature hyperparameter scaling similarity magnitudes, and summation indices $k$ traverse randomly sampled negative annotations or sequences. The dual logarithmic terms enforce mutual retrievability constraints: protein sequences should distinguish their true annotations from decoys, while annotations should identify their corresponding sequences.
\paragraph{Direct Preference Optimization (DPO) Loss}
By parameterizing human preference probabilities through the optimal policy $\pi_\theta$ rather than explicit reward modeling, we derive the Direct Preference Optimization (DPO) objective:
\begin{equation}
\resizebox{1.0\linewidth}{!}{$\displaystyle
\mathcal{L}_\text{DPO}(\pi_\theta; \pi_\text{ref}) = -\mathbb{E}_{(x, y_w, y_l)\sim \mathcal{D}} \left[ \log\sigma\left( \beta\log\frac{\pi_\theta(y_w|x)}{\pi_\text{ref}(y_w|x)} - \beta\log\frac{\pi_\theta(y_l|x)}{\pi_\text{ref}(y_l|x)} \right) \right],
$}
\end{equation}
where $x$ denotes input protein sequences, $y_w$ represents ground-truth functional annotations from Swiss-Prot, and $y_l$  corresponds to synthetic negatives. The reference policy $\pi_\text{ref}$ preserves knowledge from the supervised fine-tuned model, while the temperature parameter $\beta>0$ controls deviation from this baseline. The sigmoid function $\sigma(\cdot)$ converts log-probability differences into preference  likelihoods.

\section{Experiments}
\paragraph{Performance Evaluation in Gene Ontology Subcategories}
We conducted a comprehensive evaluation of model performance across GO subcategories. The testset sequences were stratified by these three ontological categories and evaluated using zero-shot, SFT, and DPO models. Quantitative analysis employing F1-Max and AUPR metrics revealed substantial performance disparities (Fig. \ref{fig:go_exampe_main}a). The zero-shot approach demonstrated minimal predictive capability (F1-Max less than 0.1 across all categories), while DPO consistently outperformed SFT, achieving relative F1-Max improvements of 2.7\%, 4.1\%, and 3.1\% in Biological Process (BP), Cellular Component (CC), and Molecular Function (MF) categories respectively.

\paragraph{Long-Tail Distribution Adaptation Analysis}
To investigate model robustness against label frequency imbalance, we categorized GO terms into three frequency groups: low-frequency ($<1\%$ occurrence), medium-frequency (1-10\%), and high-frequency ($>10\%$). All models were evaluated on testset (Fig. \ref{fig:go_exampe_main}b). DPO exhibited superior performance across all frequency regimes, particularly demonstrating 8.7\%, 4.9\% and 3.2\% F1-Max improvements over SFT in the low, medium and high-frequency categories. This underscores DPO model's enhanced capability in managing rare annotations through its preference optimization framework.

\paragraph{General GO Category Discriminability}
We visualize single-category GO annotations (BP/CC/MF) from Swiss-Prot-New via t-SNE. Both sequence and annotation features form distinct clusters aligned with biological categories (Fig. \ref{fig:go_exampe_main}c). DPO demonstrates clearer separation than other baselines, particularly between molecular functions and cellular components, indicating enhanced ability to distinguish functional categories.

\paragraph{Fine-Grained Ontological Relationship Learning}
To examine hierarchical relationship capture within GO categories, we selected tightly-related GO term families (e.g., enzyme regulation in MF, protein translation in BP, cytoskeleton in CC) and visualized their sequence embeddings. Fig. \ref{fig:go_exampe_main}d demonstrates that DPO-learned features preserve ontological proximity, with related terms forming distinct subclusters. This hierarchical structure awareness enables more biologically meaningful annotation predictions.

\paragraph{Ablation Study}
Our systematic ablation analysis (Tab. \ref{tab:model_ablation}) reveals critical architectural contributions to model performance. The zero-shot model shows minimal functionality, while SFT model achieves substantial improvement. The integration of LoRA adapters provides additional gains, demonstrating the effectiveness of parameter-efficient finetuning. Our DPO models significantly outperform previous baselines, where DPO model with model-predicted annotations as negatives achieves state-of-the-art performance. Notably, the contrastive learning component proves essential for its removal degrades GO F1-Max by 67.9\% compared to full SFT.

\begin{table}[htb]
  \centering
  \caption{Ablation study on the model structure.}
  
  {
  \begin{tabular}{lccc}
    \toprule
    \multirow{1}{*}{\centering \bf{Model version}}& \multirow{1}{*}{\bf{F1-Max}} & \multirow{1}{*}{\bf{Recall}}& \multirow{1}{*}{\bf{AUROC}}\\
    \midrule
    Zero Shot      & 0.0016 & 0.4687 & 0.4941 \\
    SFT w/o SAC    & 0.2419 & 0.0686 & 0.9358 \\
    SFT            & 0.7533 & 0.6031 & 0.9891 \\
    SFT LoRA       & 0.7683 & 0.6332 & 0.9915 \\
    
    \midrule
    DPO w/ msk noise& 0.7796 & 0.6192 & 0.9961 \\
    DPO w/ pred     & \bf{0.7947} & \bf{0.7027} & \bf{0.9979} \\
    
    \bottomrule
  \end{tabular}}
  
  \label{tab:model_ablation}
\end{table}

\section{Conclusion}
In this study, we present a novel framework for protein functional annotation prediction by integrating Direct Preference Optimization into a multimodal learning pipeline. 
Our method addresses annotation sparsity through two synergistic mechanisms: contrastive alignment between sequence embeddings and GO term features during supervised finetuning and direct optimization of human-curated annotation preferences via DPO, circumventing reward modeling complexities. Experimental results demonstrate enhanced discriminability across GO categories compared to conventional approaches, with latent space visualizations revealing clear separation of biological processes, molecular functions, and cellular components. While current performance is constrained by existing annotation biases in Swiss-Prot, this work establishes a paradigm for incorporating evolving functional knowledge through preference-aware learning, enabling adaptive integration of new annotation evidence without architectural modification.

\section*{Impact Statement}
This paper pioneers the integration of contrastive learning with Direct Preference Optimization to address critical challenges in protein functional annotation: annotation sparsity and cross-modal misalignment. By eliminating reward modeling dependencies and enabling direct optimization of biological preferences, this work accelerates the discovery of uncharacterized protein functions while providing a blueprint for dynamic integration of evolving functional evidence in computational biology. The methodology extends beyond annotation prediction, offering a generalizable paradigm for human-preference-aligned learning in biological sequence analysis.
\bibliography{example_paper}
\bibliographystyle{icml2025}

\newpage
\appendix
\onecolumn
\section{Model and Training Details}
\label{model_train_details}
\begin{table}[H]
\centering
\caption{Model architecture hyperparameters}
\begin{tabular}{ll}
\hline
\textbf{Parameter} & \textbf{Value} \\
\hline
Sequence Length & 512 \\
Annotation Classes & 7533 \\
Annotation Encoder Attention Head Dimension & 64 \\
Annotation Encoder Attention Heads & 8 \\
Annotation Encoder Depth & 12 \\
Annotation Encoder Hidden Dimension & 960 \\
Annotation Encoder Global Dimension & 512 \\
Annotation Predictor Dropout Rate & 0.1 \\
Annotation Predictor Residual Blocks & 2 \\
\hline
\end{tabular}
\end{table}

\begin{table}[H]
\centering
\caption{SFT hyperparameters}
\begin{tabular}{ll}
\hline
\textbf{Parameter} & \textbf{Value} \\
\hline
Batch Size per GPU & 128 \\
Base Learning Rate & 5e-5 \\
Minimum Learning Rate & 5e-7 \\
Warmup Initial Learning Rate & 5e-7 \\
Warmup Epochs & 3 \\
Finetuning Epochs & 80 \\
Learning Rate Decay Rate & 0.95 \\
\hline
\end{tabular}
\end{table}

\begin{table}[H]
\centering
\caption{DPO hyperparameters}
\begin{tabular}{ll}
\hline
\textbf{Parameter} & \textbf{Value} \\
\hline
Batch Size per GPU & 48 \\
DPO Beta & 0.1 \\
Number of Augmentations & 3-10 \\
Training Weight & 0.01-1.0 \\
DPO Loss Weight & 0.01-1.0 \\
KL Divergence Weight & 0.1-1.0 \\
NLL Loss Weight & 0.01-100 \\
Diversity Loss Weight & 1.0 \\
SAC Loss Weight & 1.0 \\
Alpha Balance Factor & 1.0 \\
Warmup Steps & 1\% of total steps \\
DPO Total Epochs & 20 \\
Base Learning Rate & 5e-5 \\
Minimum Learning Rate & 5e-7 \\
Warmup Learning Rate & 5e-7 \\
\hline
\end{tabular}
\end{table}

\section{Dataset Details}
\label{dataset_details}
\begin{table}[H]
  \centering
  \caption{Classification of GO terms by functional category and annotation frequency.}
  
  \begin{tabular}{lr}
    \toprule
    \bf{Classification} & \bf{Amount} \\
    \midrule
    \textit{Function}\\
    \hspace{0.5cm} CC (Cellular Component) & 962\\
    \hspace{0.5cm} BP (Biological Process) & 3346\\
    \hspace{0.5cm} MF (Molecular Function) & 3225\\
    \hspace{0.5cm} Total & 7533\\
    \midrule
    \textit{Frequency}\\
    \hspace{0.5cm} Low & 4120\\
    \hspace{0.5cm} Medium & 2680\\
    \hspace{0.5cm} High & 733\\
    \hspace{0.5cm} Total & 7533\\
    \bottomrule
  \end{tabular}
  
  \label{protac-godict-details}
\end{table}

\begin{table}[H]
  \centering
  \caption{Dataset sequence counts with annotation inclusion criteria: training set totals, test set for any GO category occurrence, and Swiss-Prot-New for exclusive single-category GO terms and frequency-based counts.}
  \label{tab:go-distribution}
  \begin{tabular}{llr}
    \toprule
    \textbf{Dataset} & \textbf{Classification} & \textbf{Amount} \\
    \midrule
    
    \multirow{1}{*}{Training Set} 
      & Total & 483,285 \\
      
    \midrule
    
    \multirow{5}{*}{Test Set} 
      & \textit{Function}\\
      & \hspace{0.5cm} BP & 43,740 \\
      & \hspace{0.5cm} CC & 41,225 \\
      & \hspace{0.5cm} MF & 46,770 \\
      & Total & 53,563 \\
      
    \midrule
    
      \multirow{9}{*}{Swiss-Prot-New} 
      & \textit{Frequency}\\
      & \hspace{0.5cm} Low ($\textless$1\%)& 2,501 \\
      & \hspace{0.5cm} Medium (1\%$\sim$10\%)& 15,625 \\
      & \hspace{0.5cm} High ($\textgreater$10\%)& 37,148 \\
      \cmidrule{2-3}
      & \textit{Function}\\
      & \hspace{0.5cm} BP & 385 \\
      & \hspace{0.5cm} CC & 1,829 \\
      & \hspace{0.5cm} MF & 1,970 \\
      \cmidrule{2-3}
      & Total & 37,972\\
    \bottomrule
  \end{tabular}
\end{table}

\section{Experiment Details}

\setlength\tabcolsep{2pt}
\begin{table}[H]
  \centering
  \scriptsize
  \caption{Quantitative performance metrics across GO subcategories.}
  {
    \begin{tabular}{lcccccccccccccccc}
      \toprule
      \multirow{2}{*}{\centering \bf{Model version}}& \multicolumn{5}{c}{\bf{BP}} & \multicolumn{5}{c}{\bf{CC}}& \multicolumn{5}{c}{\bf{MF}}\\
      \cmidrule(r){2-6}
      \cmidrule(r){7-11}
      \cmidrule(r){12-16}    
      &\bf{Recall} & \bf{Precision} & \bf{F1-Max} & \bf{AUROC} & \bf{AUPR} &\bf{Recall} & \bf{Precision} & \bf{F1-Max} & \bf{AUROC} & \bf{AUPR} &\bf{Recall} & \bf{Precision} & \bf{F1-Max} & \bf{AUROC} & \bf{AUPR}\\
      \midrule
      Zero Shot & 0.4599 & 0.0006 & 0.0013 & 0.4752 & 0.0006 & 0.3597 & 0.0014 & 0.0044 & 0.4430 & 0.0016 & 0.5571 & 0.0008 & 0.0019 & 0.5450 & 0.0008\\
      SFT & 0.5381 & 0.9575 & 0.7155 & 0.9875 & 0.7205 & 0.5579 & 0.8990 & 0.7124 & 0.9933 & 0.7565 & 0.7663 & 0.9591 & 0.8604 & 0.9938 & 0.8986\\
      DPO & 0.6075 & 0.9171 & 0.7345 & 0.9958 & 0.7488 & 0.6344 & 0.8742 & 0.7418 & 0.9975 & 0.7977 & 0.8329 & 0.9481 & 0.8870 & 0.9992 & 0.9294\\
      \bottomrule
    \end{tabular}
    }
  \label{tab:cross-category-metrics}
\end{table}

\begin{table}[H]
  \centering
  \caption{Quantitative performance of robustness evaluation across annotation frequency groups.}
  {
  \begin{tabular}{lccccccccc}
    \toprule
    \multirow{2}{*}{\centering \bf{Model version}}& \multicolumn{3}{c}{\bf{GO F1-Max}} & \multicolumn{3}{c}{\bf{GO Recall}}& \multicolumn{3}{c}{\bf{GO AUROC}}\\
    \cmidrule(r){2-4}
    \cmidrule(r){5-7}
    \cmidrule(r){8-10}    
    &\bf{Low} & \bf{Medium} & \bf{High} &\bf{Low} & \bf{Medium} & \bf{High} &\bf{Low} & \bf{Medium} & \bf{High} \\
    \midrule
    Zero shot & 0.0026 & 0.0017 & 0.0015 & 0.4843 & 0.4609 & 0.4549 & 0.4956 & 0.4877 & 0.4909 \\
    SFT       & 0.4224 & 0.5710 & 0.6238 & 0.2197 & 0.4275 & 0.5110 & 0.9464 & 0.9639 & 0.9651 \\
    DPO       & 0.4591 & 0.5992 & 0.6439 & 0.3114 & 0.4985 & 0.5620 & 0.9708 & 0.9783 & 0.9773\\
    \bottomrule
  \end{tabular}}
  
  \label{tab:frequency-robustness}
\end{table}

\section{Additional Experiment Results}
\label{add_exp_results}

\begin{figure}[htb]
    \centering
    \includegraphics[width=1\linewidth]{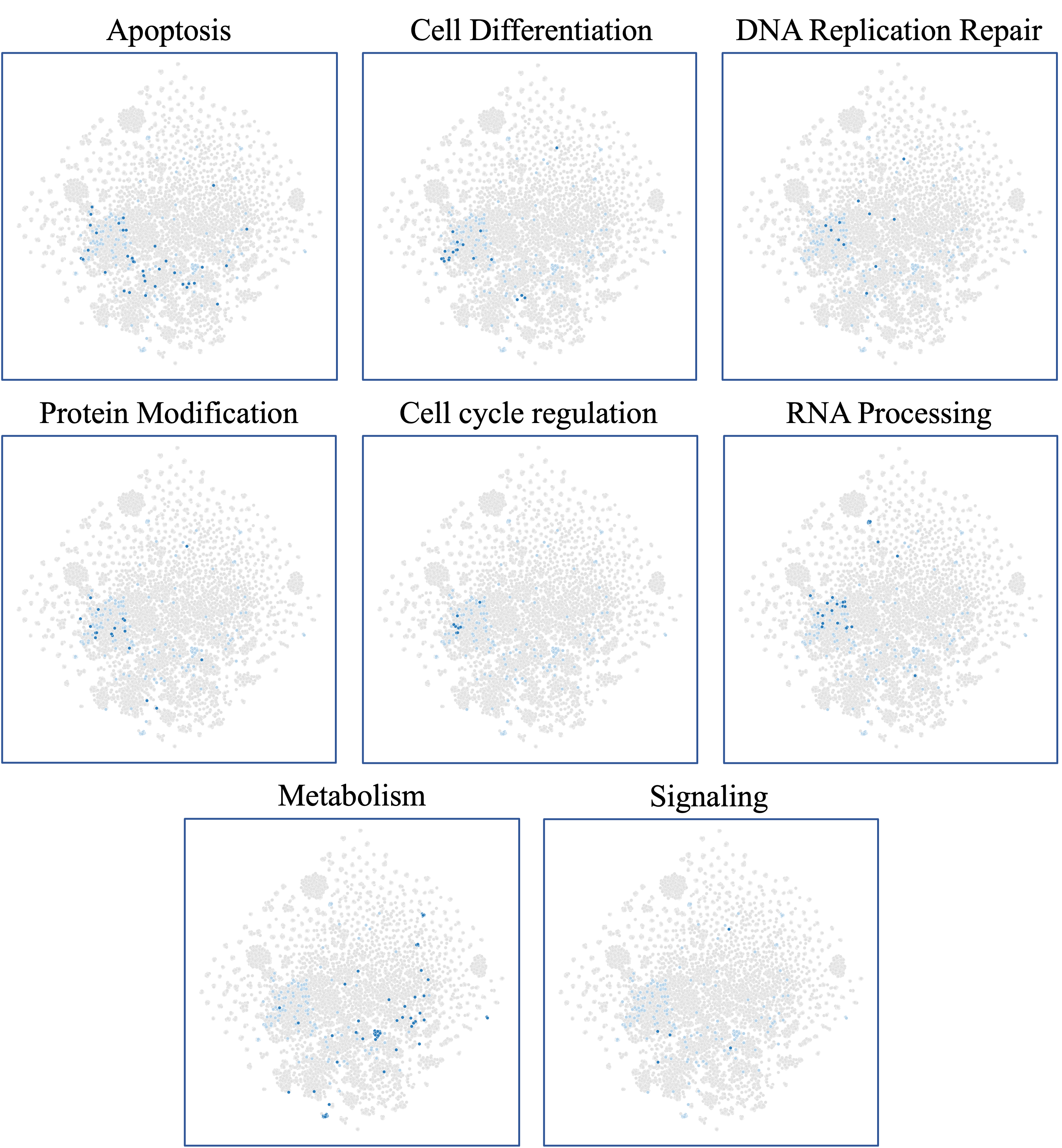}
    
    \caption{Additional results of sequences with biological process related GOs in fine-grained ontological relationship learning task.}
    
    \label{fig:go_example_bp}
\end{figure}

\begin{figure}[htb]
    \centering
    \includegraphics[width=1\linewidth]{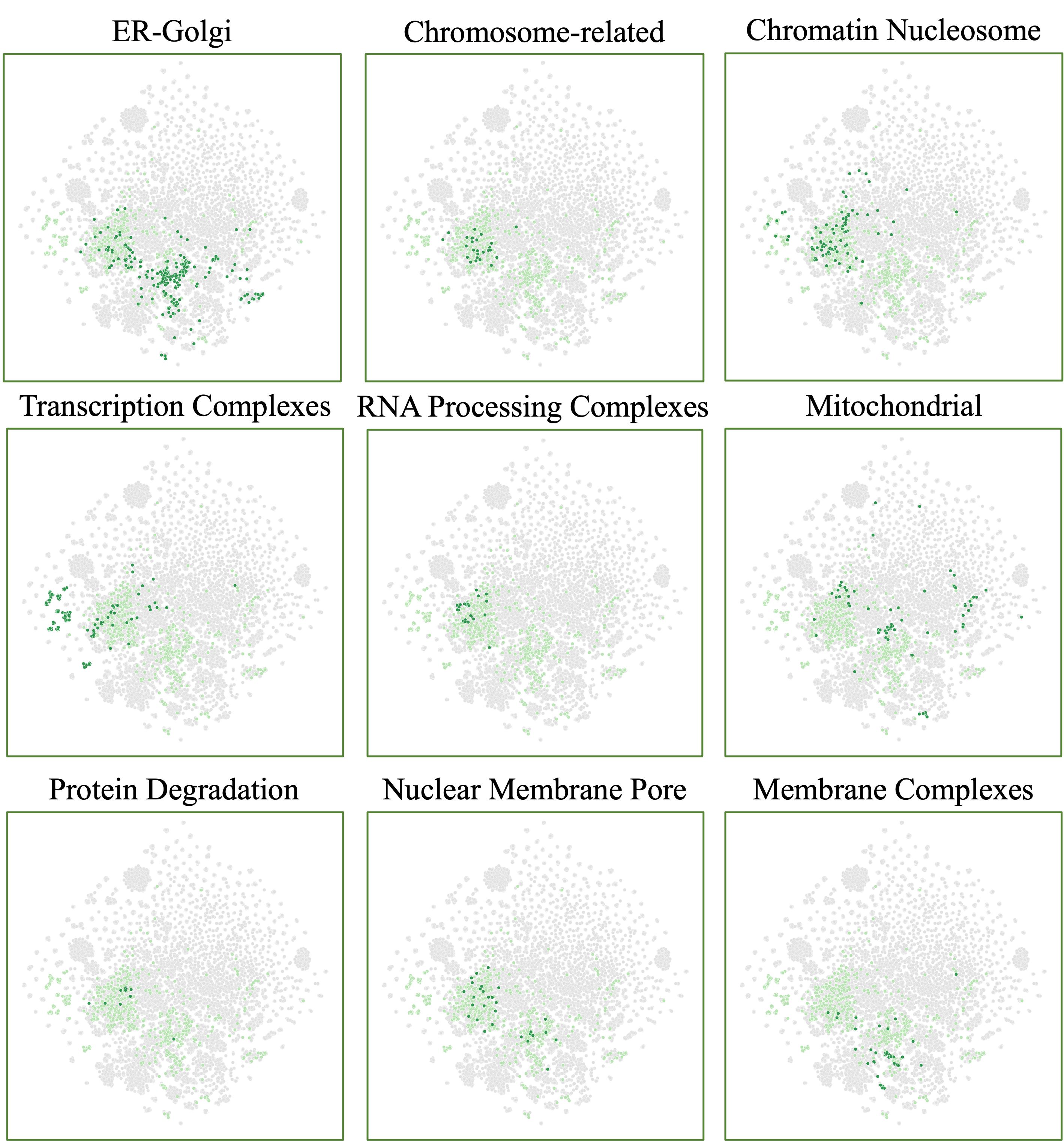}
    
    \caption{Additional results of sequences with cellular component related GOs in fine-grained ontological relationship learning task.}
    
    \label{fig:go_example_cc}
\end{figure}

\begin{figure}[htb]
    \centering
    \includegraphics[width=1\linewidth]{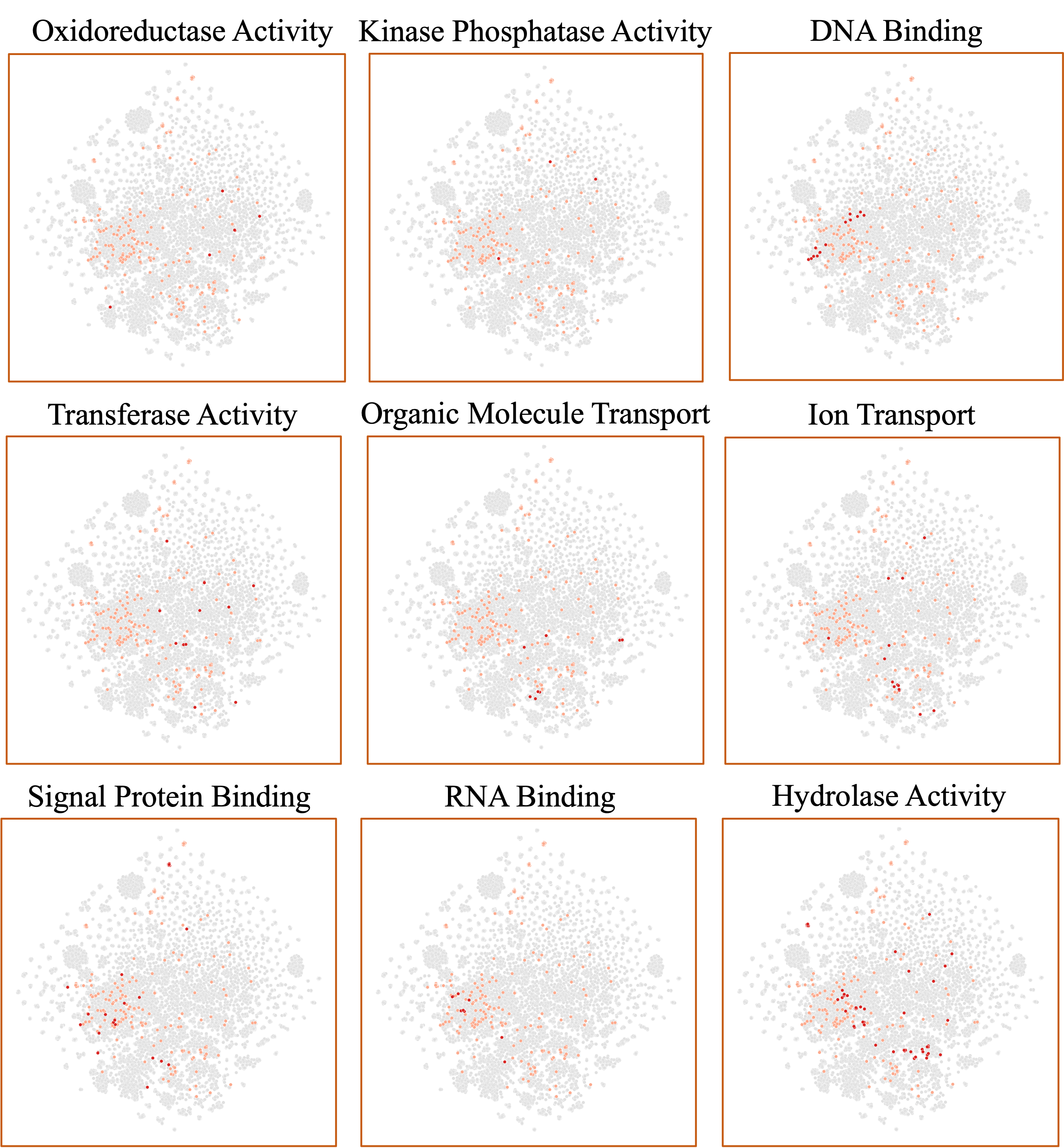}
    
    \caption{Additional results of sequences with molecular function related GOs in fine-grained ontological relationship learning task.}
    
    \label{fig:go_example_mf}
\end{figure}

\begin{table}[H]
  \centering
  \scriptsize
  \caption{Additional GOs of biological process.}
  
  \begin{tabularx}{\textwidth}{l|c|c|c|X}
    \toprule
    \bf{Category} &\bf{Subcategory} & \bf{Sequence amount} & \bf{GO ID} & \bf{Term}\\
    \midrule
    \multirow{44}{*}{Biological Process} & \multirow{3}{*}{Apoptosis} & \multirow{3}{*}{56} & GO:0006915 & Apoptotic process\\
    & & & GO:2001235 & Positive regulation of apoptotic signaling pathway\\
    & & & GO:0043027 & Cysteine-type endopeptidase inhibitor activity involved in apoptotic process\\
    
    \cmidrule{2-5}
    
    & \multirow{5}{*}{Cell Cycle Regulation} & \multirow{5}{*}{11} & GO:0007050 & Regulation of cell cycle\\
    & & & GO:2000045 & Regulation of G1/S transition of mitotic cell cycle\\
    & & & GO:0007049 & Cell cycle\\
    & & & GO:0070192 & Chromosome organization involved in meiosis\\
    & & & GO:0007142 & Male meiosis II\\
    
    \cmidrule{2-5}

    & \multirow{6}{*}{Cell Differentiation} & \multirow{6}{*}{30} & GO:0048741 & Skeletal muscle fiber development\\
    & & & GO:0021954 & Central nervous system neuron development\\
    & & & GO:0048513 & Animal organ development\\
    & & & GO:0048666 & Neuron development\\
    & & & GO:0045595 & Regulation of cell differentiation\\
    & & & GO:0060173 & Limb development\\
    
    \cmidrule{2-5}

    & \multirow{5}{*}{DNA Replication Repair} & \multirow{5}{*}{32} & GO:0006271 & DNA strand elongation involved in DNA replication\\
    & & & GO:0032297 & Negative regulation of DNA-templated DNA replication initiation\\
    & & & GO:0071897 & DNA biosynthetic process\\
    & & & GO:0006290 & Pyrimidine dimer repair\\
    & & & GO:0006267 & Pre-replicative complex assembly involved in nuclear cell cycle DNA replication\\
    
    \cmidrule{2-5}

    & \multirow{5}{*}{Metabolism} & \multirow{5}{*}{300} & GO:0006739 & NADP metabolic process\\
    & & & GO:0006644 & Phospholipid metabolic process\\
    & & & GO:0016042 & Lipid catabolic process\\
    & & & GO:0019563 & Glycerol catabolic process\\
    & & & GO:0006083 & Acetate metabolic process\\
    
    \cmidrule{2-5}

    & \multirow{6}{*}{Protein Modification} & \multirow{6}{*}{24} & GO:0031398 & Positive regulation of protein ubiquitination\\
    & & & GO:0035871 & Protein K11-linked deubiquitination\\
    & & & GO:0071569 & Protein ufmylation\\
    & & & GO:0001934 & Positive regulation of protein phosphorylation\\
    & & & GO:0035307 & Positive regulation of protein dephosphorylation\\
    & & & GO:0031146 & SCF-dependent proteasomal ubiquitin-dependent protein catabolic process\\
    
    \cmidrule{2-5}

    & \multirow{4}{*}{Protein Translation} & \multirow{4}{*}{35} & GO:0002183 & Cytoplasmic translational initiation\\
    & & & GO:0006415 & Translational termination\\
    & & & GO:0045900 & Negative regulation of translational elongation\\
    & & & GO:0002182 & Cytoplasmic translational elongation\\
    
    \cmidrule{2-5}

    & \multirow{5}{*}{RNA Processing} & \multirow{5}{*}{79} & GO:0031167 & rRNA methylation\\
    & & & GO:0000288 & Nuclear-transcribed mRNA catabolic process, deadenylation-dependent decay\\
    & & & GO:0000967 & rRNA 5'-end processing\\
    & & & GO:0006406 & mRNA export from nucleus\\
    & & & GO:0000956 & Nuclear-transcribed mRNA catabolic process\\
    
    \cmidrule{2-5}

    & \multirow{5}{*}{Signaling} & \multirow{5}{*}{8} & GO:0038166 & Angiotensin-activated signaling pathway\\
    & & & GO:0007259 & Cell surface receptor signaling pathway via JAK-STAT\\
    & & & GO:0033209 & Tumor necrosis factor-mediated signaling pathway\\
    & & & GO:0030520 & Estrogen receptor signaling pathway\\
    & & & GO:0010469 & Regulation of signaling receptor activity\\
    
    \bottomrule
  \end{tabularx}
  
  \label{tab:bp-go-seqs}
\end{table}

\begin{table}[H]
  \centering
  \scriptsize
  \caption{Additional GOs of cellular component.}
  
  \begin{tabularx}{\textwidth}{l|c|c|c|X}
    \toprule
    \bf{Category} &\bf{Subcategory} & \bf{Sequence amount} & \bf{GO ID} & \bf{Term}\\
    \midrule
    \multirow{52}{*}{Cellular Component} & \multirow{5}{*}{Chromatin Nucleosome} & \multirow{5}{*}{145} & GO:0005721 & Pericentric heterochromatin\\
    & & & GO:0000779 & Condensed chromosome, centromeric region\\
    & & & GO:0000792 & Heterochromatin\\
    & & & GO:0031519 & PcG protein complex\\
    & & & GO:0005694 & Chromosome\\
    
    \cmidrule{2-5}
    
    & \multirow{5}{*}{Chromosome-related} & \multirow{5}{*}{35} & GO:0000922 & Spindle pole\\
    & & & GO:0000940 & Outer kinetochore\\
    & & & GO:1990879 & CST complex\\
    & & & GO:0000930 & Gamma-tubulin complex\\
    & & & GO:0035371 & Microtubule plus-end\\
    
    \cmidrule{2-5}

    & \multirow{7}{*}{Cytoskeleton} & \multirow{7}{*}{136} & GO:0005925 & Focal adhesion\\
    & & & GO:0005912 & Adherens junction\\
    & & & GO:0070161 & Anchoring junction\\
    & & & GO:0097431 & Mitotic spindle pole\\
    & & & GO:0036064 & Ciliary basal body\\
    & & & GO:0036157 & Outer dynein arm\\
    & & & GO:0001534 & Radial spoke\\
    
    \cmidrule{2-5}

    & \multirow{4}{*}{ER-Golgi} & \multirow{4}{*}{592} & GO:0005789 & Endoplasmic reticulum membrane\\
    & & & GO:0090158 & Endoplasmic reticulum membrane organization\\
    & & & GO:0005784 & Sec61 translocon complex\\
    & & & GO:0005802 & Trans-Golgi network\\
    
    \cmidrule{2-5}

    & \multirow{8}{*}{Membrane Complexes} & \multirow{8}{*}{66} & GO:0009897 & External side of plasma membrane\\
    & & & GO:0031241 & Periplasmic side of cell outer membrane\\
    & & & GO:0098982 & GABA-ergic synapse\\
    & & & GO:0045211 & Postsynaptic membrane\\
    & & & GO:0005921 & Gap junction\\
    & & & GO:0005922 & Connexin complex\\
    & & & GO:0034707 & Chloride channel complex\\
    & & & GO:0030867 & Rough endoplasmic reticulum membrane\\
    
    \cmidrule{2-5}

    & \multirow{6}{*}{Mitochondrial} & \multirow{6}{*}{146} & GO:0005759 & Mitochondrial matrix\\
    & & & GO:0005744 & Mitochondrial inner membrane presequence translocase complex\\
    & & & GO:0030964 & NADH dehydrogenase complex\\
    & & & GO:0070469 & Respiratory chain\\
    & & & GO:0042645 & Mitochondrial nucleoid\\
    & & & GO:0005761 & Mitochondrial ribosome\\
    
    \cmidrule{2-5}

    & \multirow{3}{*}{Nuclear Membrane Pore} & \multirow{3}{*}{52} & GO:0031965 & Nuclear membrane\\
    & & & GO:0071765 & Nuclear inner membrane organization\\
    & & & GO:0031080 & Nuclear pore complex\\
    
    \cmidrule{2-5}

    & \multirow{3}{*}{Protein Degradation} & \multirow{3}{*}{16} & GO:0000151 & Ubiquitin ligase complex\\
    & & & GO:0019005 & SCF ubiquitin ligase complex\\
    & & & GO:0031464 & Cul4-RING E3 ubiquitin ligase complex\\
    
    \cmidrule{2-5}

    & \multirow{6}{*}{RNA Processing Complexes} & \multirow{6}{*}{48} & GO:0005681 & Spliceosomal complex\\
    & & & GO:0071006 & U2-type catalytic step 1 spliceosome\\
    & & & GO:0071007 & U2-type catalytic step 2 spliceosome\\
    & & & GO:0089701 & U2 snRNP\\
    & & & GO:0005685 & U1 snRNP\\
    & & & GO:0005849 & mRNA cleavage factor complex\\
        
    \cmidrule{2-5}

    & \multirow{10}{*}{Transcription Complexes} & \multirow{10}{*}{731} & GO:0005666 & RNA polymerase III complex\\
    & & & GO:0000428 & DNA-directed RNA polymerase complex\\
    & & & GO:0016580 & Sin3 complex\\
    & & & GO:0016592 & Mediator complex\\
    & & & GO:0030880 & RNA polymerase complex\\
    & & & GO:0005673 & Transcription factor TFIIE complex\\
    & & & GO:0016586 & RSC-type complex\\
    & & & GO:0032783 & Super elongation complex\\
    & & & GO:0090575 & RNA polymerase II transcription factor complex\\
    & & & GO:0000118 & Histone deacetylase complex\\
    
    \bottomrule
  \end{tabularx}
  
  \label{tab:cc-go-seqs}
\end{table}

\begin{table}[H]
  \centering
  \scriptsize
  \caption{Additional GOs of molecular function.}
  
  \begin{tabularx}{\textwidth}{l|c|c|c|X}
    \toprule
    \bf{Category} & \bf{Subcategory} & \bf{Sequence amount} & \bf{GO ID} & \bf{Term}\\
    \midrule
    \multirow{60}{*}{Molecular Function} & \multirow{6}{*}{Hydrolase Activity} & \multirow{6}{*}{141} & GO:0016798 & Hydrolase activity, acting on glycosyl bonds\\
    & & & GO:0070004 & Cysteine-type exopeptidase activity\\
    & & & GO:0008234 & Cysteine-type peptidase activity\\
    & & & GO:0004045 & Aminoacyl-tRNA hydrolase activity\\
    & & & GO:0016920 & Pyroglutamyl-peptidase activity\\
    & & & GO:0004843 & Thiol-dependent deubiquitinase activity\\
    
    \cmidrule{2-5}
    
    & \multirow{6}{*}{Transferase Activity} & \multirow{6}{*}{54} & GO:0016765 & Transferase activity, transferring alkyl or aryl groups\\
    & & & GO:0008318 & Protein prenyltransferase activity\\
    & & & GO:0004057 & Arginyl-tRNA--protein transferase activity\\
    & & & GO:0015019 & Heparan-alpha-glucosaminide N-acetyltransferase activity\\
    & & & GO:0008791 & Arginine N-succinyltransferase activity\\
    & & & GO:0047173 & Phosphatidylcholine-retinol O-acyltransferase activity\\
    
    \cmidrule{2-5}
    
    & \multirow{8}{*}{Oxidoreductase Activity} & \multirow{8}{*}{9} & GO:0016714 & Oxidoreductase activity, acting on paired donors\\
    & & & GO:0004174 & Electron-transferring-flavoprotein dehydrogenase activity\\
    & & & GO:0004471 & Malate dehydrogenase (decarboxylating) (NAD+) activity\\
    & & & GO:0047111 & Formate dehydrogenase (cytochrome-c-553) activity\\
    & & & GO:0004665 & Prephenate dehydrogenase (NADP+) activity\\
    & & & GO:0046553 & D-malate dehydrogenase (decarboxylating) (NAD+) activity\\
    & & & GO:0003834 & Beta-carotene 15,15'-dioxygenase activity\\
    & & & GO:0016630 & Protochlorophyllide reductase activity\\
    
    \cmidrule{2-5}
    
    & \multirow{5}{*}{Kinase Phosphatase Activity} & \multirow{5}{*}{18} & GO:0106311 & Protein serine/threonine kinase activity\\
    & & & GO:0004797 & Thymidine kinase activity\\
    & & & GO:0004703 & G protein-coupled receptor kinase activity\\
    & & & GO:0008673 & 2-dehydro-3-deoxygluconokinase activity\\
    & & & GO:0004331 & Fructose-2,6-bisphosphate 2-phosphatase activity\\
    
    \cmidrule{2-5}
    
    & \multirow{6}{*}{Ion Transport} & \multirow{6}{*}{33} & GO:0015087 & Cobalt ion transmembrane transporter activity\\
    & & & GO:0008324 & Cation transmembrane transporter activity\\
    & & & GO:0005221 & Intracellularly cyclic nucleotide-activated monoatomic cation channel activity\\
    & & & GO:0005223 & Intracellularly cGMP-activated cation channel activity\\
    & & & GO:0008308 & Voltage-gated monoatomic anion channel activity\\
    & & & GO:0015444 & P-type magnesium transporter activity\\
    
    \cmidrule{2-5}
    
    & \multirow{7}{*}{Organic Molecule Transport} & \multirow{7}{*}{13} & GO:0015187 & Glycine transmembrane transporter activity\\
    & & & GO:0015181 & L-arginine transmembrane transporter activity\\
    & & & GO:0005324 & Long-chain fatty acid transmembrane transporter activity\\
    & & & GO:0015221 & Lipopolysaccharide transmembrane transporter activity\\
    & & & GO:0090482 & Vitamin transmembrane transporter activity\\
    & & & GO:0015655 & Alanine:sodium symporter activity\\
    & & & GO:0015189 & L-lysine transmembrane transporter activity\\
    
    \cmidrule{2-5}
    
    & \multirow{5}{*}{DNA Binding} & \multirow{5}{*}{30} & GO:1990837 & Sequence-specific double-stranded DNA binding\\
    & & & GO:0000986 & Cis-regulatory region sequence-specific DNA binding\\
    & & & GO:0000404 & Heteroduplex DNA loop binding\\
    & & & GO:0043138 & 3'-5' DNA helicase activity\\
    & & & GO:1990970 & Trans-activation response element binding\\
    
    \cmidrule{2-5}
    
    & \multirow{5}{*}{RNA Binding} & \multirow{5}{*}{15} & GO:0008143 & Poly(A) binding\\
    & & & GO:0045131 & Pre-mRNA branch point binding\\
    & & & GO:0001070 & RNA binding transcription factor activity\\
    & & & GO:0030619 & U1 snRNA binding\\
    & & & GO:0033897 & Ribonuclease T2 activity\\
    
    \cmidrule{2-5}
    
    & \multirow{6}{*}{Signal Protein Binding} & \multirow{6}{*}{75} & GO:0005132 & Type I interferon receptor binding\\
    & & & GO:0005164 & Tumor necrosis factor receptor binding\\
    & & & GO:0008190 & Eukaryotic initiation factor 4E binding\\
    & & & GO:0031072 & Heat shock protein binding\\
    & & & GO:0008013 & Beta-catenin binding\\
    & & & GO:0044325 & Ion channel binding\\
    
    \cmidrule{2-5}
    
    & \multirow{6}{*}{Enzyme Regulation} & \multirow{6}{*}{355} & GO:0005096 & GTPase activator activity\\
    & & & GO:0004864 & Protein phosphatase inhibitor activity\\
    & & & GO:0030234 & Enzyme regulator activity\\
    & & & GO:0043022 & Ribosome binding\\
    & & & GO:0010521 & Telomerase inhibitor activity\\
    & & & GO:0030337 & DNA polymerase processivity factor activity\\
    
    \bottomrule
  \end{tabularx}
  \label{tab:mf-go-seqs}
\end{table}
\end{document}